# Variational U-Net with Local Alignment for Joint Tumor Extraction and Registration (VALOR-Net) of Breast MRI Data Acquired at Two Different Field Strengths

Muhammad Shahkar Khan[1], MS*; Haider Ali[1,2], PhD*; Laura Villazan Garcia[2], BS;   Noor Badshah[3], PhD; Siegfried Trattnig[4,5], PhD; Florian Schwarzhans[2], PhD;   Ramona Woitek[2,6], PhD; Olgica Zaric[2,5], PhD†

[1]Department of Mathematics, University of Peshawar, Peshawar, Khyber Pakhtunkhwa, Pakistan.

[2]Research Group for Medical Image Analysis and Artificial Intelligence, Danube Private University, Krems, Austria.

[3]Department of Basic Sciences and Islamiyat, University of Engineering & Technology, Peshawar, Pakistan.

[4]High-field MR Centre, Medical University of Vienna, Vienna, Austria

[5]Institute for Musculoskeletal imaging, Karl Landsteiner Society, St. Pölten, Austria

[6]Department of Radiology, School of Clinical Medicine, University of Cambridge, Cambridge Biomedical Campus, Cambridge, UK

*These authors contributed equally to this work

†Corresponding author: Olgica Zaric, PhD

 Research Centre for Medical Imaging and Image Analysis (MIAAI)

Viktor Kaplan-Straße 2, 2700 Wiener Neustadt, Austria

Danube Private University (DPU), Steiner Landstr. 124, 3500 Krems-Stein, Austria

E-Mail: olgica.zaric@dp-uni.ac.at

**Abstract:**

**Background:** Multiparametric breast MRI data might improve tumor diagnostics, characterization, and treatment planning. Accurate alignment and delineation of images acquired at different field strengths such as 3T and 7T, remain challenging research tasks.

**Purpose:** To address alignment challenges and enable consistent tumor segmentation across different MRI field strengths.

**Study type:** Retrospective.

**Subjects:** Nine female subjects with breast tumors were involved: six histologically proven invasive ductal carcinomas (IDC) and three fibroadenomas.

**Field strength/sequence:** Imaging was performed at 3T and 7T scanners using post-contrast T1-weighted three-dimensional time-resolved angiography with stochastic trajectories (TWIST) sequence.

**Assessments:** The method's performance for joint image registration and tumor segmentation was evaluated using several quantitative metrics, including signal-to-noise ratio (PSNR), structural similarity index (SSIM), normalized cross-correlation (NCC), Dice coefficient, F1 score, and relative sum of squared differences (rel SSD).

**Statistical tests:** The Pearson correlation coefficient was used to test the relationship between the registration and segmentation metrics.

**Results:** When calculated for each subject individually, the PSNR was in a range from 27.5 to 34.5 dB, and the SSIM was from 82.6 to 92.8%. The model achieved an NCC from 96.4 to 99.3% and a Dice coefficient of 62.9 to 95.3%. The F1 score was between 55.4 and 93.2% and the rel SSD was in the range of 2.0 and 7.5%. The segmentation metrics Dice and F1 Score are highly correlated (0.995), while a moderate correlation between NCC and SSIM (0.681) was found for registration.

**Data conclusion:** Initial results demonstrate that the proposed method may be feasible in providing joint tumor segmentation and registration of MRI data acquired at different field strengths.

# 1 Introduction

Magnetic resonance imaging (MRI) has become a cornerstone in the diagnostic evaluation of breast tumors due to its ability to provide exceptional soft tissue contrast and detail. Increased signal-to-noise ratio (SNR) and superior spatial resolution at 7T enable the visualization of intricate anatomical details and insight into biochemical changes of pathological tissue often undetectable with conventional 3T scanners [1, 2]. Moreover, aligning 3T data to the 7T reference allows for effective fusion of morphological or semi-functional images generated on 3T with X-nuclei images acqiured at 7T, creating a comprehensive tumor profile [3-9]. This fusion enhances clinical decision-making by integrating diverse tumor characteristics, such as tumor metabolic activity, in addition to morphological data.

The image registration involves the estimation of geometric transformations by optimizing similarity measures. While methods like the sum of squared differences (SSD) are common, they assume Gaussian noise and are vulnerable to outliers, which can lead to inaccuracies in noisy medical data [10]. Regularization techniques, such as diffusion and total variation, help guide solutions but struggle with medical challenges such as organ boundary discontinuities and complex deformations due to respiration or pathological variations [6, 11-13]. Higher-order regularizations improve performance but complicate optimization and reduce computational efficiency [6]. Recently, U-Net, a convolutional neural network used for medical image segmentation [14], offers an implicit regularization through its encoder-decoder architecture, skip connections, and multi-scale feature aggregation. These features help preserve fine details and broader context, ensuring smooth and robust outputs for medical image registration [15]. Interestingly, Atlas-ISTN is a deep-

VALOR-Net for Segmentation and Registration of the Breast MRI Data

learning framework that jointly learns segmentation and registration, constructing a population-derived atlas while ensuring topologically consistent segmentations. However, its reliance on large, labeled training datasets [12, 16] presents a challenge in clinical scenarios where annotated data may be limited [5]. However, the authors have recently proposed a multi-modality registration model in [5], which also requires an extensive training dataset. Recently, a hybrid deep-learning model combining CNNs with stochastic gradients demonstrated improved breast cancer detection accuracy through dataset integration, although it remains limited by computational complexity, extensive parameter tuning [16, 17], and challenges in scalability with high-dimensional data [3]. Deep-learning (DL) has revolutionized numerous fields, including computer vision and natural language processing. However, it faces critical challenges such as reliance on large labeled datasets, high computational costs, and the absence of standardized guidelines for model selection and training. Moreover, DL models often lack interpretability, making their predictions difficult to understand and limiting their application in sensitive domains like medical imaging. These challenges are especially pronounced in scenarios where data is scarce, computational resources are constrained, or transparency is essential [2, 18-20].

Medical image analysis traditionally employs model-based [20-27], machine-learning-based [15, 11, 25, 28, 29], or hybrid approaches [11], each offering distinct strengths and limitations. Model-based techniques are known for their robustness but often require meticulous parameter tuning, particularly due to the inclusion of regularization terms [17, 18, 23]. For instance, parameter optimization for each individual slice is necessary in some cases, which is impractical and time-consuming. These variational models heavily depend on precise parameter tuning tailored to the specific characteristics of the data, posing a significant challenge for widespread implementation. Conversely, deep-learning-based approaches, such as those utilizing convolutional neural networks (CNNs), face hurdles stemming from the high variability of medical imaging data across patients. This variability demands a diverse and representative dataset for effective training, which is often difficult to obtain, particularly in specialized medical domains [1]. Addressing these challenges is crucial for advancing the application of DL in medical imaging and improving its accessibility in resource-limited settings. Amid these challenges, Deep Image Prior (DIP) emerges as a promising alternative. Unlike traditional DL methods, DIP leverages the inherent structure of convolutional neural networks to solve image reconstruction problems without requiring extensive external training data [28]. By relying solely on the network's architecture, DIP effectively reduces data dependency, lowers computational costs, and enhances interpretability, making it particularly valuable for medical imaging applications [16]. In addition, DIP demonstrates robustness in addressing inverse problems, offering an efficient and reliable framework that mitigates the limitations of conventional DL approaches [17, 27]. This combination of advantages positions DIP as a transformative tool in the field of medical image analysis. The existing method [11] improves upon previous models [7-10] in three key aspects: it employs a relaxed formulation for segmentation; selectively uses geometric markers from the target image (T) to segment both the target (T) and reference (R); and utilizes a deep image prior-based approach for more accurate joint segmentation and registration [30]. As an initial framework, it is currently incapable of segmenting normal-sized or small tumors, with suboptimal registration performance. Moreover, the use of two independent U-Net architectures results in increased computational time and memory consumption without providing any mutual guidance between the segmentation and registration processes.

To overcome the limitations of previously developed segmentation and registration methods, we aimed to develop an algorithm with the following key characteristics:

1.  joint segmentation and registration: unlike traditional approaches that treat segmentation and registration as separate tasks, we aimed to develop a method that integrates them into a single framework;

2.  support for multi-modality imaging data: the goal is to establish a framework that effectively handles data acquired on different MRI systems, broadening applicability across diverse clinical settings;

3.  simultaneous processing of multiple slices: the proposed method can process multiple slices concurrently, improving computational efficiency and scalability compared to models that handle slices individually;

4.  tumor size independence: a novel selective segmentation term should be able to provide a robust performance regardless of tumor size, addressing a common limitation of existing segmentation models;

5.  enhanced registration and reconstruction quality: through a novel local alignment mechanism, we aim to achieve superior registration accuracy and reconstruction quality while preserving local features;

6.  data-independent test-train framework: this would incorporate variational models in its loss functions to reduce reliance on large labeled datasets, making it efficient for scenarios like 7T breast MRI with limited data availability.

## 2 Methodology

### 2.1 Subjects

Ten female subjects with breast tumors were involved. One subject was excluded due low data quality. This study was approved by the institutional review board, and written, informed consent was obtained from all participants. This work was done in collaboration with the XXX (ethics approval number: XXX, cooperation agreement number: XXX). All the patients were examined on the 3T scanner (Prisma Siemens Healthineers, Erlangen, Germany) and the 7T scanner (Magnetom, Siemens Healthineers, Erlangen, Germany). The subjects were scanned in a prone position using the proton ($^1$H) bilateral breast coils (Stark, Erlangen Germany).

## 2.2 MRI Sequences

Imaging was performed using T1-weighted three-dimensional time-resolved angiography with stochastic trajectories (TWIST) sequence optimized for both 3T and 7T scanners, providing high temporal resolution of 14 seconds and high spatial resolution. For the 3T scanner, the echo time (TE) and repetition time (TR) were set to 2.84 ms and 6.01 ms, respectively. The acquisition time was nine minutes, and the spatial resolution was $1.1 \times 1.1 \times 1.1$ mm$^3$. For the 7T scanner, the TE and TR were set to 2.5 ms and 4.75 ms, respectively. The acquisition time remained nine minutes, and the spatial resolution was $0.7 \times x0.7 \times 0.7$ mm$^3$.

## 2.3 Preprocessing for Registration and Segmentation

To prepare 3T MRI breast volumes for registration with 7T volumes and facilitate tumor segmentation, a streamlined preprocessing pipeline was applied to both the target (3T, $T$) and reference (7T, $R$) MRI volumes (Figure 1). Tumor-containing slices were selected based on clinical interest, with specific slice numbers and cropping margins adjusted according to anatomical variation in each volume. Peripheral regions were cropped to focus on areas of diagnostic relevance. Next, intensity normalization and histogram matching were applied to align the intensity distributions of $T$ and $R$, reducing modality-specific contrast differences. CLAHE (Contrast Limited Adaptive Histogram Equalization) was used to enhance the visibility of low-contrast regions, especially in tumor areas. Each preprocessed slice was resized to a standardized resolution (320×320 pixels) to ensure consistency in registration and segmentation tasks. For cases with tumors that exhibited extremely low contrast relative to surrounding tissue, segmentation was performed at the original resolution to maintain high segmentation performances.

## 2.4 Data Postprocessing and Evaluations

Segmentation and registration metrics were calculated for all tumor slices to evaluate performance across the entire volume. As ground truth, we used the manual segmenetation done by a biomedical engineer and supervised by a medical imaging physicist (XXX and XXX) using ITK-SNAP software (version 4.0.0). The slice-specific metrics were aggregated and visualized as bar diagrams. The postprocessing procedure is divided into key stages as outlined below. The information on software and sripts used in this work may be found here: https://github.com/WadudWali/Joint-Segmentation-and-Registration.

## 2.5 Statistical Analysis

A statistical analysis using the Pearson correlation coefficient matrix was performed to assess the model's performance in simultaneously performing segmentation and registration where -1 indicates a perfectly negative linear correlation between two variables, 0 indicates no linear correlation between two variable and 1 indicates a perfectly positive linear correlation between two variables. This analysis was performed using statistical software R version 4.4.2 (The R Foundation for Statistical Computing), where the computed metrics for segmentation, including F1 score and Dice coefficient, along with registration metrics such as normalized cross correlation coefficient (NCC), structural similarity index (SSIM), and peak signal-to-noise ratio (PSNR), were utilized to evaluate the relationship between these different measures of model performance.

## 2.6 Novel Variational Loss Function

The proposed model is defined as a comprehensive framework tailored for the joint segmentation and registration of MRI breast images obtained from 7T and 3T systems. This innovative approach leverages variational principles to simultaneously optimize tumor delineation and spatial alignment between images of different magnetic field strengths. To achieve this, we designed a robust loss function that integrates local and global features, enhancing the model's capability to accurately identify and segment breast tumors while ensuring precise registration across modalities. The loss function is given by:

$$F(\theta, u)^{loss} = \mu \int_{\Omega} D(x) \cdot |\theta(x) - T(x)| \, d\Omega + \lambda_1 \int_{\Omega} \Phi(T, a_1)\theta(x) \, d\Omega +$$

$$\lambda_2 \int_{\Omega} \Phi(R, c_1)\theta(x + u) \, d\Omega$$

$$+ \beta_1 \int_{\Omega} |T(x + u) - R(x)|^2 \, d\Omega + \beta_2 \int_{\Omega} [(T_s(x + u) - T(x + u) -$$

$$R_s(x) + R(x))^2] \, d\Omega \tag{3.1}$$

where $\theta(x)$ is the segmentation tensor at a spatial location $x$ within the image domain $\Omega$, $u$ is the deformation field that warps the template image, $T(x)$ is the template image, and $R(x)$ is the reference image, $T_s(x + u)$ and $R_s(x)$ are the smoothed versions of the warped template and reference images, respectively, and $\mu, \lambda_1, \lambda_2, \beta_1, \beta_2$ are weight parameters controlling the contributions of each term. In our study, we use a fixed set of parameters: $\lambda_1 = 2$, $\lambda_2 = 0.1$, $\beta_1 = 50$, $\beta_2 = 50$, and $\mu = 30$.

Term-by-term explanation of the loss function:

1. First term:

$$\mu \int_\Omega D(x) \cdot |\theta(x) - T(x)| \, d\Omega$$

This term incorporates the geometrical shape prior to selective constraint $D(x)$ defined as user inputs which represents the geodesic distance that penalizes objects away from the user-marked region $M$. $D(x)$ enforces alignment between the segmentation $\theta(x)$ and the template image $T(x)$, favoring segmentation boundaries close to regions indicated by the user.

The geodesic distance $D(x)$ is computed as:

$$D(x) = \frac{D_0(x)}{\|D_0(x)\|_{L^\infty}}$$

where $D_0(x)$ is calculated based on the gradient of the template image $\nabla T(x)$ and edge map information, enforcing that the segmented object aligns with prior knowledge or user input.

2. Second and Third terms:

$$\lambda_1 \int_\Omega \Phi(T, a_1)\theta(x) \, d\Omega + \lambda_2 \int_\Omega \Phi(R, c_1)\theta(x + u) \, d\Omega$$

These terms provide statistical shape priors to guide the segmentation process. The function $\Phi(f, a)$ is defined as:

$$\Phi(f, a) = \mu_1(f, a) - \mu_2(f, a)$$

where $\mu_1(f, a) = (f - a)^2$, and $\mu_2(f, a)$ modulates the segmentation based on the distance of the intensity value $f$ to the average intensity $a$. The parameters $a_1$ and $c_1$ are the statistical mean intensities of the region of interest in the template image $T$ and the reference image $R$, respectively.

The second term applies this prior to the segmentation of the template image, while the third term applies it to the warped template image $T(x + u)$, ensuring that the deformation field $u$ also respects the statistical priors in the reference image.

3. Fourth term:

$$\beta_1 \int_\Omega |T(x + u) - R(x)|^2 \, d\Omega$$

This term ensures *global alignment* between the deformed template image $T(x + u)$ and the reference image $R(x)$. By minimizing the squared difference between these images, the model encourages an overall correspondence between the two, helping to refine the deformation field $u$ so that $T(x + u)$ aligns with $R(x)$.

4. Fifth term:

$$\beta_2 \int_\Omega \left[ (T_s(x + u) - T(x + u) - R_s(x) + R(x))^2 \right] d\Omega$$

This term enhances *local feature alignment* by comparing the smoothed versions of the template and reference images. $T_s(x + u)$ and $R_s(x)$ are the smoothed images of the warped template and reference, respectively, where smoothing helps reduce noise and emphasizes larger structures.

### 2.7 U-Net Architecture for Joint Tumor Segmentation and Registration

The proposed U-Net architecture, designed for joint tumor segmentation and registration of MRI 7T and 3T images, leverages two interconnected U-Net models with distinct roles: one dedicated to segmentation and the other to registration. This architecture is illustrated in Figure 4.

a)  U-Net 1: Tumor Segmentation

The first U-Net model, referred to as U-Net 1, is responsible for tumor segmentation. It receives preprocessed MRI slices as input and processes them through a series of down-sampling and up-sampling blocks. The Down-Sampling Blocks reduce the spatial dimensions of the input while increasing feature channels. Each block consists of two convolutional layers followed by PReLU activation, Batch Normalization, and Dropout for regularization. A Bottleneck Layer processes the lowest-resolution feature maps and enables feature aggregation. The Up-Sampling Blocks progressively restore spatial resolution and refine the segmentation by concatenating up-sampled features with the corresponding down-sampled features, followed by convolutional layers, PReLU, Batch Normalization, and Dropout. The output of U-Net 1 is a segmentation map, generated by a convolutional layer with a sigmoid activation function, representing the segmented tumor region. Following the tumor segmentation, a Spatial Transformer Layer is employed to align the template (3T) image with the reference (7T) image. This layer applies learned transformations based on the estimated deformation field to warp the template image, aligning it with the reference image and facilitating accurate tumor localization.

b) U-Net 2: Joint Registration and Segmentation Refinement

The second U-Net model, U-Net 2, integrates the outputs from U-Net 1 and performs joint registration and further segmentation refinement. This model receives the segmentation map, the warped template image, and other relevant input channels for improved performance. Similar to U-Net 1, U-Net 2 includes Down-Sampling Blocks for feature extraction and Up-Sampling Blocks for restoring spatial resolution and refining the segmentation. The final output of U-Net 2 is a Deformation Field, which facilitates further image registration by warping both the template and reference images, and refining the segmentation maps. This deformation field is concatenated with the template image, reference image, and other relevant tensors to produce the final output. For the network architecture, the number of input filters is set to 32 and the learning rates for both stages are fixed at $lr1 = 0.001$ and $lr2 = 0.001$, ensuring stable convergence during optimization.

c) The Final Outputs

The final outputs of the proposed architecture include several key components essential for accurate tumor localization and analysis. The deformation field is used to warp both the images and segmentation maps, aligning the 3T and 7T MRI volumes. This ensures precise tumor localization across different MRI field strengths. The unmatched images represent the template image (3T), which is warped to align with the reference image (7T), facilitating the comparison between the two modalities. In addition, the segmentation maps are generated to identify tumor regions in both the 3T and 7T MRI volumes, providing accurate tumor delineation. Finally, the warped segmentation maps are transformed according to the deformation field, enabling further analysis of the segmented tumor regions.

**3 Results**

Nine subjects (mean age±standard deviation) 56±12 years old with breast tumors (six malignant and three benign) were included in the study. Both segmentation and registration metrics were recorded at intervals of 50 and 200 epochs during training to monitor the model's progress. Results were visualized (Fig. 2-10) and tabulated (Table 1) to facilitate a detailed analysis of individual tumor slices within the 3T breast volumes. The resulting subplots, displayed in Figure 11 (a-i), illustrate the distribution of metrics for each patient individually. The maximum PSNR achieved was 34.54 dB, indicating high-quality images with minimal noise. The SSIM reached a maximum of 92.76%, reflecting a strong structural similarity between registered images, and signifying an accurate alignment. The NCC achieved a maximum value of 99.24%, demonstrating the model's robust capability to align images across different MRI field strengths. For tumor segmentation, VALOR-Net reached a maximum Dice coefficient of 95.32%, indicating high delineation capabilities of the tumor area. The F1 score, which accounts for both precision and recall, was 93.14%, reflecting accurate tumor segmentation with a low false-positive rate. The minimum rel SSD observed was 1.06%, highlighting minimal geometric distortion during the registration process. The joint model appears to perform well, particularly in segmentation, with good agreement in registration (Table 2). The weak negative correlation between registration and segmentation performance suggests that the trade-off is minimal and the model performs both tasks effectively.

## 4 Discussion

In this paper, we have proposed and demonstrated the efficacy of a novel framework that integrates joint segmentation and registration for 3T and 7T breast MRI images.

The results, presented in Figures 2-11, provide both qualitative and quantitative insights into the model's performance. Each figure contains sample slices from nine patients, illustrating segmentation and registration results for both imaging modalities. The model's ability to accurately capture small and complex tumor shapes in challenging scenarios is noteworthy.

### 4.1 Qualitative and Quantitative Performance

The qualitative results demonstrate that the registration successfully aligns 3T breast MRI slices to 7T slices, producing anatomically consistent images. Notably, the transformed 3T images often revealed additional anatomical details, including those seen on the 7T images, which could enhance clinical interpretation. This is evident in Figures 2-10, where the transformed 3T slices on 7T consistently achieved a visual quality comparable to 7T, with minimal discrepancies. Quantitatively, registration metrics, such as PSNR, achieved values above 90% and around 32, respectively, indicating high fidelity in alignment.

For segmentation, the model successfully delineated tumors across various challenging scenarios. For example, in subject 1 (Fig. 2a), the tumor was poorly detectable due to its small size and poorly defined margins. However, the model accurately segmented this tumor, achieving above 80% for segmentation metrics (Fig. 2b, Fig. 11a). Similarly, in subject 2, the tumor exhibited an irregular appearance in the 7T slice and an oval shape in the corresponding 3T slice (Fig. 3a). Despite these differences, the model segmented the tumor effectively in both modalities, as corroborated by segmentation metrics (Jaccard, Dice, and F1 score) exceeding 80% (Fig. 3b, Fig. 11b).

Subject 3 presented another unique challenge: a non-mass tumor appearance in the 3T slice, which had a lower signal ennhacement in the corresponding 7T slice (Fig. 4a). The proposed model successfully captured the tumor while maintaining a segmentation accuracy above 80%, showcasing its robustness in handling complex cases (Fig. 4b, Fig. 11c)). The qualitative results demonstrate that the registration successfully aligned 3T breast MRI slices to 7T slices, producing anatomically consistent images.

For subject 4, the transformation of the 3T slice onto the 7T space provided an anatomical view with a higher signal intensity in the tumor compared to the given 7T slice, with a moderate signal enhancement. The transformed 3T efficiently reduced these inomogeneities, enhancing the visibility of the tissue features (Fig. 5a). The quantitative metrics, including PSNR (approximately 32), NCC, SSIM, SSD, and NGF, all indicated strong registration performance with good readings on the scale, confirming the quality of the transformation (Fig. 5b). For segmentation, the model successfully captured tumors across all modalities (7T, transformed 3T, and given 3T). The tumors were clearly delineated in each sample slice, and the quantitative evaluation in the adjacent plot shows that all three segmentation metrics (Jaccard, Dice, and F1 score) were consistently above 80%. This further underscores the robustness of the model in accurately segmenting tumors even in challenging cases (Fig. 11d) .

Subject 5 also benefited from the model's registration and segmentation capabilities. The tumor size in this case was relatively large, which led to higher segmentation metrics. The tumors were well-segmented in 7T, transformed 3T, and given 3T slices (Fig. 6a, Fig. 6b). The Jaccard, Dice, and F1 score metrics were even higher compared to subject 4, reflecting the model's sensitivity to larger tumors, which is consistent with the improved performance seen with larger tumor sizes (Fig. 11e).

For subject 6, the registration results again demonstrated a satisfying performance. The transformed 3T slice on the 7T space showed an increased signal intensity, addressing areas where the 7T image was burdenered by signal loss, most probably due to inadequate coil sensitivity profiles (Fig. 7a). The quantitative metrics in this case—PSNR above 32, NCC and SSIM above 90%, and a smaller rel SSD—indicate a high-quality transformation (Fig. 7b). For segmentation, the proposed model effectively captured tumors across all modalities. The segmentation metrics again showed values above 80% for Jaccard, Dice, and the F1 score, highlighting the model's robustness in tumor detection and delineation (Fig. 11f).

The qualitative results demonstrate that the registration successfully aligned 3T breast MRI slices to 7T slices, producing anatomically consistent images. For subject 7, the transformation of the 3T slice onto the 7T space produced an optimal signal intensity of the anatomical view, capturing details that were either missing or faded in the original 7T slice (Fig. 8a). The quantitative metrics, such as PSNR (above 30%, and the registration accuracy close to 95%), reflect a strong registration performance, with good readings on the scale, confirming the quality of the transformation (Fig. 8b). For segmentation, the model performed very well in capturing tumors across all three modalities: 7T, transformed 3T, and given 3T slices. The tumors were clearly delineated in each sample slice, and the quantitative evaluation in the adjacent plot shows that all three segmentation metrics (Jaccard, Dice, and F1 score) were consistently above 80%. This indicates that the model robustly handles tumor segmentation even in the presence of small tumors or tumors with low conspicuity (Fig. 11g).

In subject 8 the registration results again demonstrated high-quality performance. The transformed 3T slice provides an image with substantially higher image quality compared to the original 7T slice (Fig. 9a). The quantitative metrics in this case—registration accuracy near 100%, PSNR above 30%—suggest an excellent transformation quality (Fig. 9b). For segmentation, although the tumor size in this case was small, the model successfully captured the tumor across all modalities. The tumor was visible in the 7T slice, the transformed 3T slice, and the given 3T slice. The quantitative evaluation of segmentation metrics (Jaccard, Dice, and F1 score) showed values above 80%, even with the small tumor size, which highlights the robustness of the model in capturing tumors despite challenges posed by size (Fig. 11h).

VALOR-Net for Segmentation and Registration of the Breast MRI Data

The qualitative results for subject 9 demonstrate that the registration successfully aligned 3T breast MRI slices to 7T slices, producing anatomically consistent images. The registration reconstruction of the 3T image on the 7T background shows a feature-rich view. This alignment leads to a more complete representation of the anatomy (Fig. 10a). The quantitative registration metrics for this subject were above 90%, with a PSNR near 30, indicating excellent registration accuracy (Fig. 10b). In addition, the tumor is clearly delineated across all three slices, with segmentation metrics showing values near 90%, underscoring the model's ability to accurately segment tumors even in complex imaging scenarios (Fig. 11i).

## 4.2 Aggregate Performance Across Tumor Slices

Fig. 11 presents a comprehensive view of the model's aggregate performance across all tumor slices for each patient, with subplots showing bar diagrams for registration and segmentation metrics. These aggregate results provide a clearer understanding of the model's overall capabilities across various tumor sizes and imaging modalities.

For subject 1 (Fig. 11a), the registration metrics show a consistently high performance between 90% and 100%, while the segmentation metrics for all tumor slices in the 3T image are above 75%. This indicates the model's robustness even in challenging scenarios where the tumor is small.

Similarly, for subject 2 (Fig. 11b), the registration metrics are close to 100%, and the segmentation metrics remain around 75%, demonstrating strong registration performance but relatively lower segmentation accuracy due to the tumor's heterogeneous appearance.

For subject 3 (Fig. 11c), the aggregate registration performance is near 100%, and the segmentation metrics are close to 90%. This high segmentation performance suggests that the model is highly effective in accurately segmenting tumors, even with complex morphology.

The aggregate performance for subjects 4, 5, and 6 shows similar trends. For subject 4 (Fig. 11d), the registration performance is close to 100%, while the segmentation accuracy is around 80%. For subject 5 (Fig. 11e) and subject 6 (Fig. 11f), the registration metrics again show a good performance, with segmentation metrics consistently above 75%.

The performance for subjects 7, 8, and 9 is high (Fig. 11g, Fig. 11h, Fig. 11i). For subject 8, the registration metrics are close to 100%, and the segmentation metrics are consistently above 90%, indicating high performance in both registration and tumor segmentation. This demonstrates the model's ability to handle more complex cases with smaller or subtle tumors.

The main limitation of this study is the low number of subjects included in the study. The image qualty of 7T data was not excellent; the regional signal drops were notable and were most probably due to the relatively large field of views used for breast imaging and the presence of external field inhomogeneity. The second reason could be that we used a four-channel double-sided array coil with a poor sensitivity profile at 7T, which could lead to local signal loss that can be reduced with an algorithm to correct the inhomogeneity of the transmit field. With improved quality of 7T data, our method might work better, which could be investigated in future studies.

## 5 Conclusion

Through enhanced registration and segmentation achieved by a local alignment mechanism, we have demonstrated substantial improvement in the performance of simultaneous image registration and segmentation. Moreover, our data-independent test-train framework has the potential to overcome the challenge of limited labeled data, which is often encountered in advanced imaging techniques such as 7T MRI. Overall, our method offers a comprehensive and robust solution applicable to both clinical practice and future research.

**Table 1**: The registration and segmentation measures (mean±standard deviation (SD)) for patients calculated for 20 slices

| Subject No. | NCC | SSIM | PSNR | rel SSD | Dice Coefficient | F1 Score |
|---|---|---|---|---|---|---|
| 1 | 0.988±0.002 | 0.89±0.02 | 30.46±0.80 | 0.035±0.007 | 0.63±0.10 | 0.55±0.10 |
| 2 | 0.972±0.001 | 0.87±0.02 | 27.51±0.80 | 0.075±0.007 | 0.78±0.10 | 0.70±0.10 |
| 3 | 0.964±0.004 | 0.84±0.02 | 31.16±1.00 | 0.020±0.007 | 0.87±0.10 | 0.83±0.04 |
| 4 | 0.974±0.001 | 0.91±0.01 | 31.05±0.60 | 0.023±0.002 | 0.82±0.04 | 0.76±0.05 |
| 5 | 0.989±0.003 | 0.88±0.06 | 28.82±1.00 | 0.036±0.010 | 0.87±0.05 | 0.81±0.06 |
| 6 | 0.978±0.003 | 0.88±0.01 | 31.18±0.60 | 0.064±0.010 | 0.77±0.06 | 0.70±0.07 |
| 7 | 0.965±0.003 | 0.83±0.01 | 27.49±0.40 | 0.048±0.004 | 0.78±0.02 | 0.70±0.03 |
| 8 | 0.989±0.003 | 0.91±0.02 | 30.35±1.00 | 0.032±0.008 | 0.87±0.07 | 0.82±0.09 |
| 9 | 0.979±0.004 | 0.85±0.07 | 29.04±0.90 | 0.047±0.010 | 0.91±0.02 | 0.87±0.03 |

*NCC- Normalized Cross-Correlation; SSIM-Structural Similarity Index; PSNR-Peak Signal-to-Noise ratio; rel SSD-Relative Sum of Squared Differences.

VALOR-Net for Segmentation and Registration of the Breast MRI Data

**Table 2:** The correlation matrix shows the correlation coefficients between the registration and segmentation metrics for nine patient data sets

| Metric | NCC | SSIM | PSNR | relSSD | Dice Coefficient | F1 score |
|---|---|---|---|---|---|---|
| NCC | 1.000 | 0.681 | 0.174 | -0.093 | -0.105 | -0.087 |
| SSIM | 0.681 | 1.000 | 0.463 | -0.197 | -0.167 | -0.154 |
| PSNR | 0.174 | 0.463 | 1.000 | -0.550 | -0.042 | 0.032 |
| rel SSD | -0.093 | -0.197 | -0.550 | 1.000 | -0.225 | -0.284 |
| Dice Coefficient | -0.105 | -0.167 | -0.042 | -0.225 | 1.000 | 0.995 |
| F1 score | -0.087 | -0.154 | 0.032 | -0.284 | 0.995 | 1.000 |

*NCC- Normalized Cross-Correlation; SSIM-Structural Similarity Index; PSNR-Peak Signal-to-Noise ratio; rel SSD-Relative Sum of Squared Differences.

**Figure Legends**

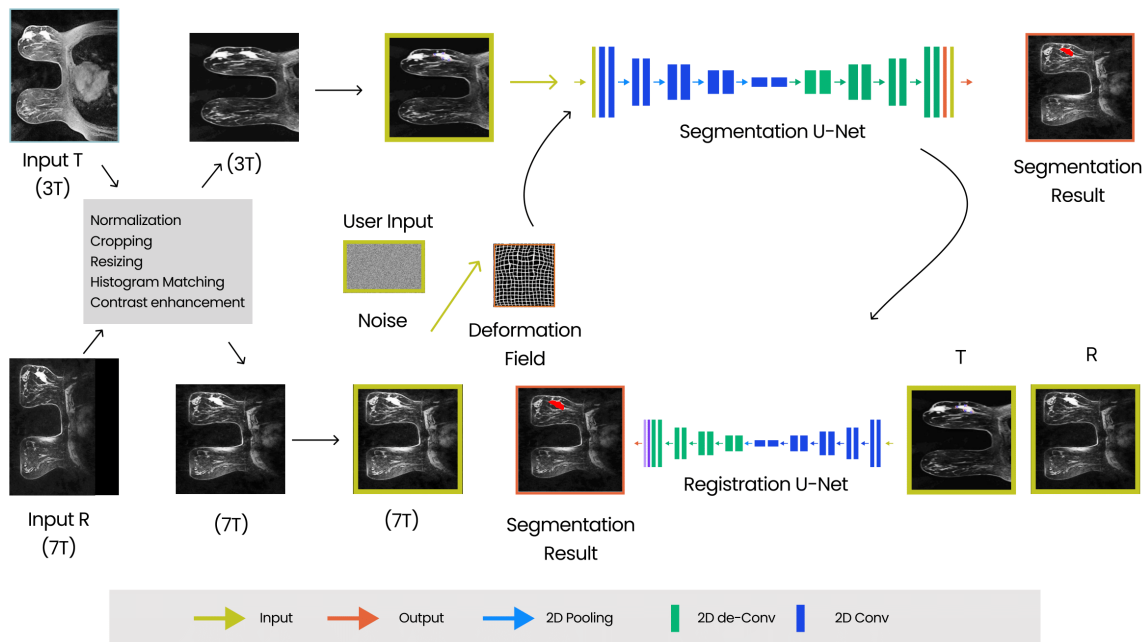

**Figure 1:** Preprocessing, postprocessing, and U-Net segmentation pipeline:

This figure outlines the methodology for preprocessing 3T MRI breast volumes, postprocessing tumor segmentation, and applying the U-Net pipeline.

**Preprocessing**: Tumor slices were selected based on clinical significance. Peripheral regions were cropped to highlight diagnostic areas. Intensity normalization and histogram matching were applied to address modality-specific contrasts. CLAHE was used to enhance low-contrast regions, particularly tumors. Slices were resized to 320x320 pixels for consistency, except when segmentation was performed at the original resolution for low-contrast tumors.

**Postprocessing**: Refinement of segmentation masks involved smoothing tumor boundaries, noise reduction, and edge sharpening via morphological operations.

**U-Net Segmentation**:

**U-Net 1**: Segmented tumors and generated segmentation masks from preprocessed slices.

**U-Net 2**: Performed deformable image registration by estimating the deformation field to align 3T and 7T volumes and warped images for precise tumor localization.

**Inputs and Outputs**: Inputs: Preprocessed 3T and 7T MRI slices, random noise. Outputs: Segmentation masks (U-Net 1) and deformation fields for image registration (U-Net 2).

VALOR-Net for Segmentation and Registration of the Breast MRI Data

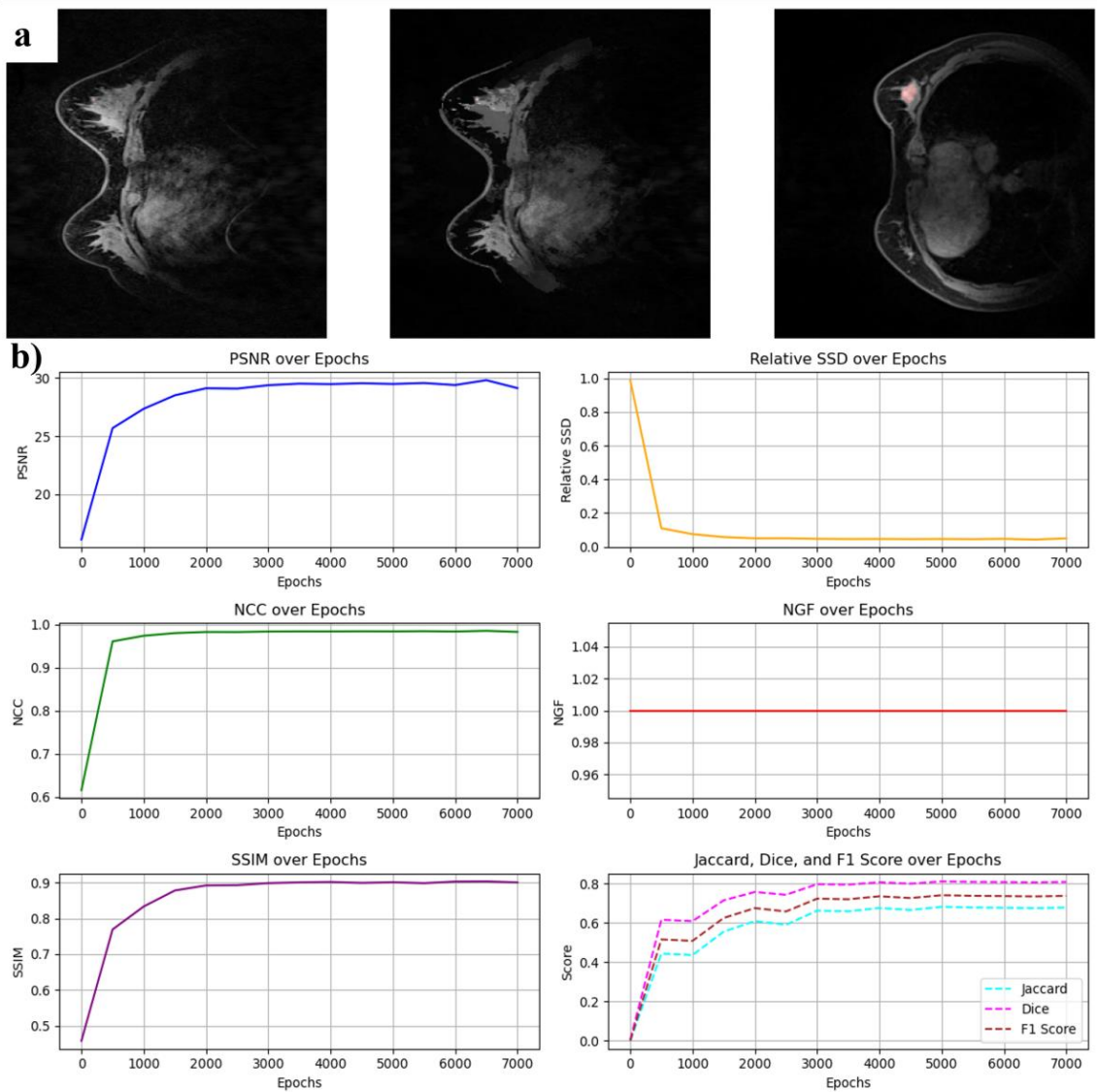

**Figure 2**: Tumor segmentation and registration results for subject 1. A 34-year-old female subject with fibroadenoma in her left breast. a) Sample slice for subject 1. From left to right: original 7T MRI image, 3T MR image transformed into the 7T space while preserving original intensities, and the original 3T MRI image. b) Quantitative metrics are presented as: Peak Signal-to-Noise Ratio (PSNR) – blue line, Normalized Cross-Correlation (NCC) – green line, Structural Similarity Index (SSIM) – violet line, Relative Sum of Squared Differences (rel SSD) – yellow line, Normalized Gradient Field (NGF) – red line, Dice coefficient – dotted pink line, F1 score – dotted dark red line, and Jacard score – dotted magenta line.

VALOR-Net for Segmentation and Registration of the Breast MRI Data

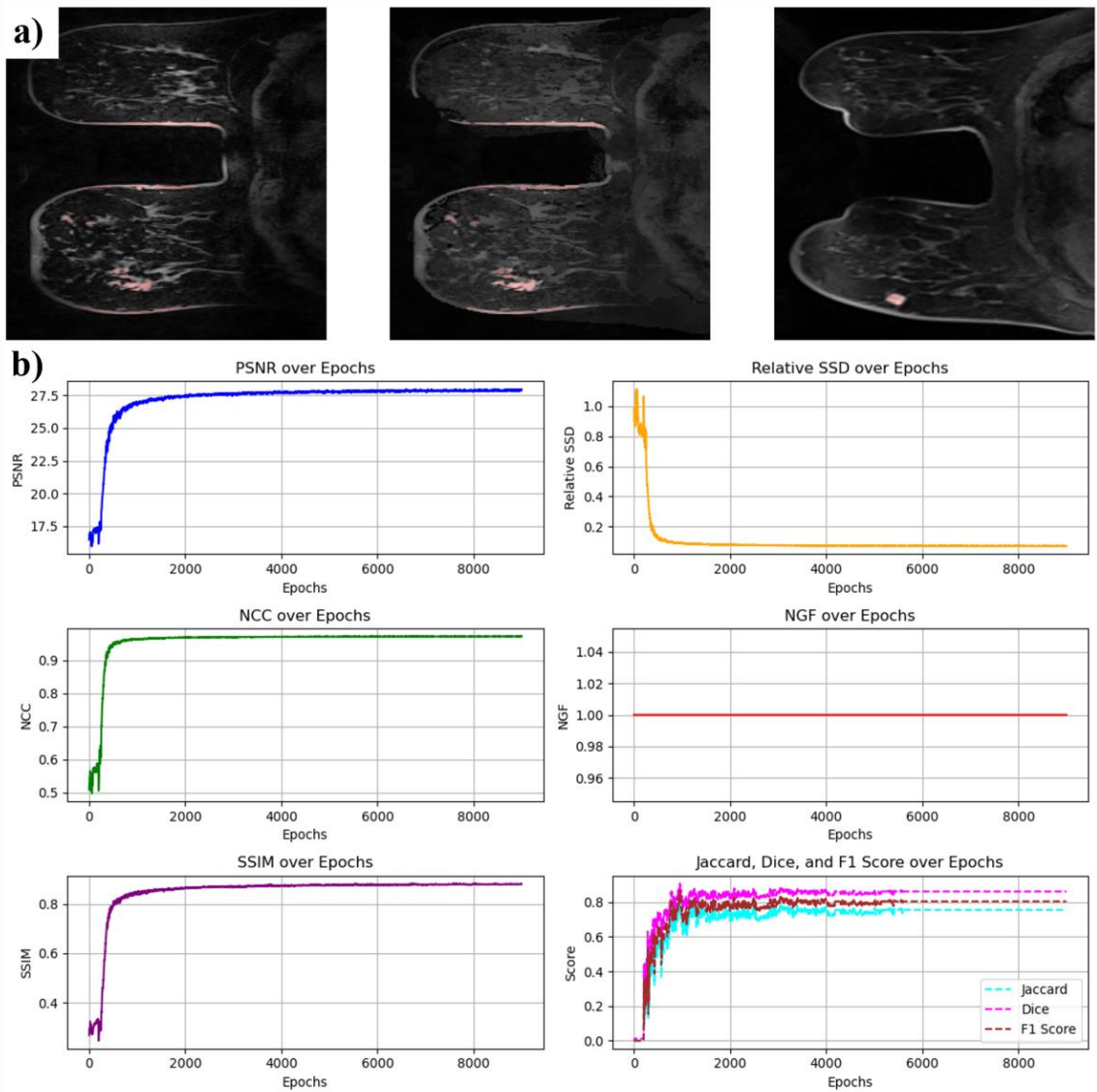

**Figure 3**: Tumor segmentation and registration results for subject 2: a) A 62-year-old female subject with invasive ductal carcinoma in her right breast. 3T MR image (left) shows heterogeneous tumor with irregular margins; 7T image demonstrates oval tumor shape (right) and transformed 3T image (middle).  b) Quantitative metrics are presented as: Peak Signal-to-Noise Ratio (PSNR) – blue line, Normalized Cross-Correlation (NCC) – green line, Structural Similarity Index (SSIM) – violet line, Relative Sum of Squared Differences (rel SSD) – yellow line, Normalized Gradient Field (NGF) – red line, Dice coefficient – dotted pink line, F1 score – dotted dark red line, and Jacard score – dotted magenta line.

VALOR-Net for Segmentation and Registration of the Breast MRI Data

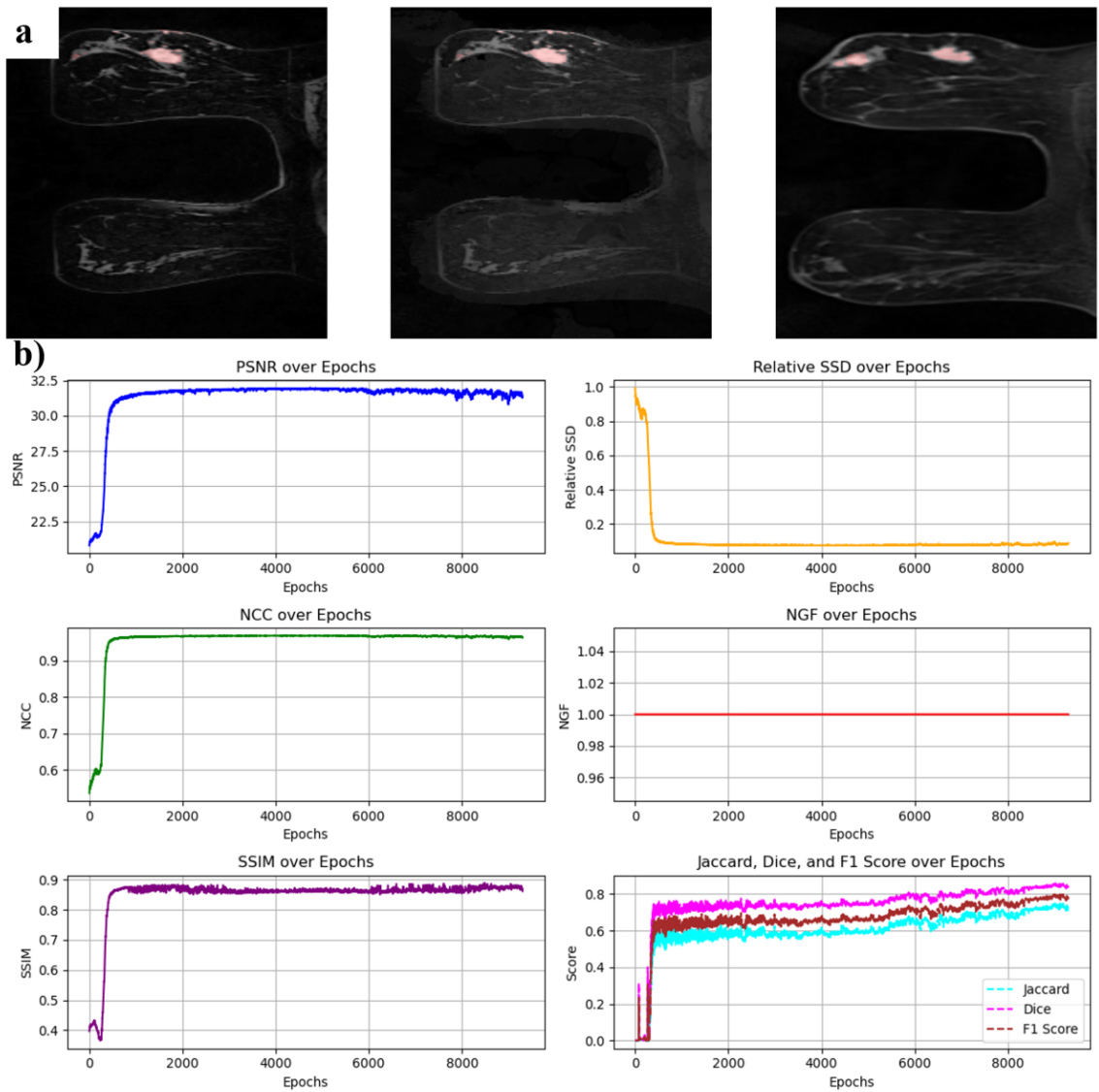

**Figure 4**: Tumor segmentation and registration results for subject 3: a) A 68-year-old female subject with invasive ductal carcinoma (IDC) in her left breast. 3T MR image (left) shows multicentric tumor detectable on 7T image (right) and on transformed 3T image (middle).   b) Quantitative metrics are presented as: Peak Signal-to-Noise Ratio (PSNR) – blue line, Normalized Cross-Correlation (NCC) – green line, Structural Similarity Index (SSIM) – violet line, Relative Sum of Squared Differences (rel SSD) – yellow line, Normalized Gradient Field (NGF) – red line, Dice coefficient – dotted pink line, F1 score – dotted dark red line, and Jacard score – dotted magenta line.

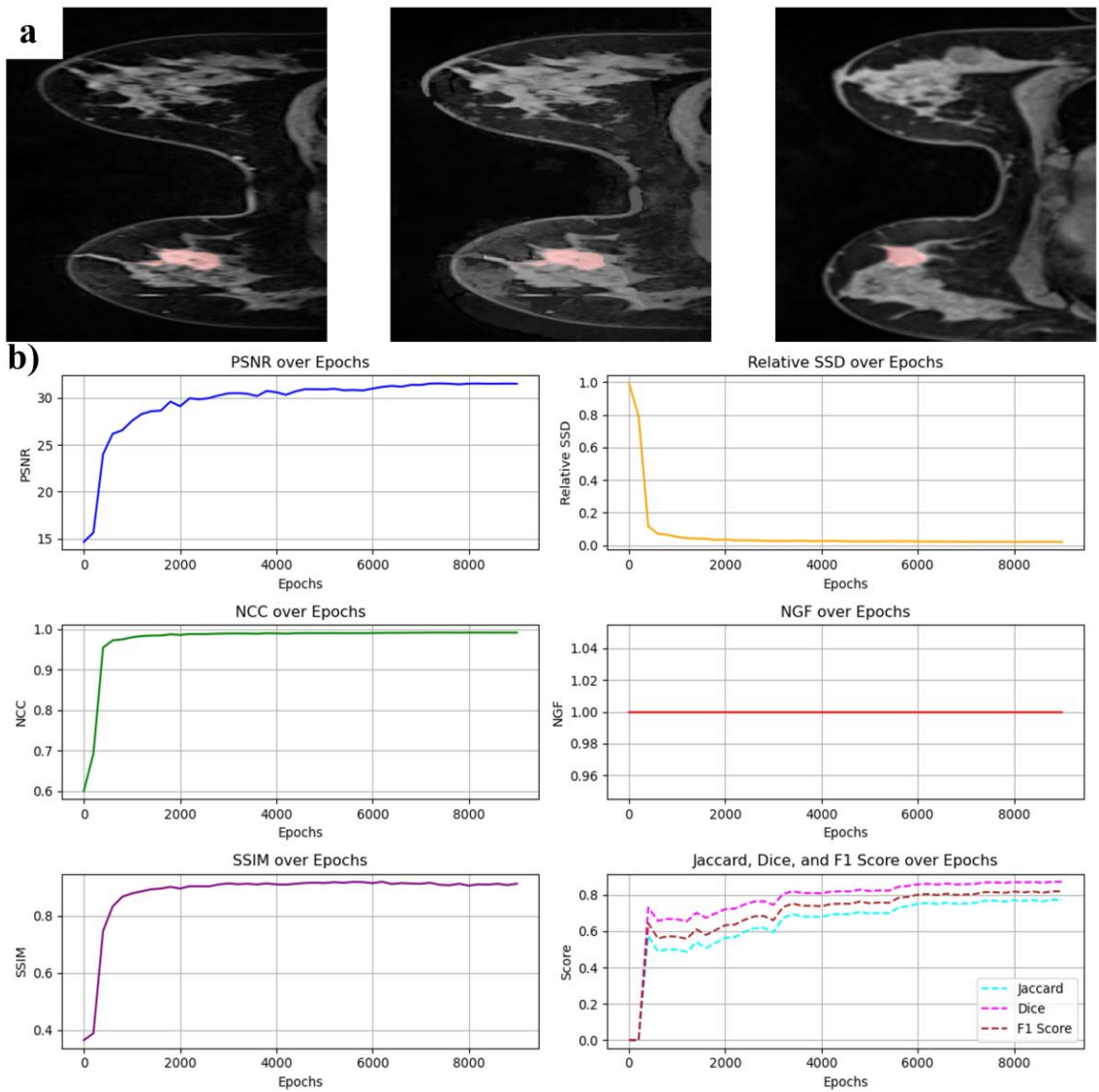

**Figure 5**: Tumor segmentation and registration results for subject 4: a) A sample slice of a 52-year-old female subject with invasive ductal carcinoma (IDC) in her left breast. The images show a tumor with irregular margins on both scanners and b) Quantitative metrics are presented as: Peak Signal-to-Noise Ratio (PSNR) – blue line, Normalized Cross-Correlation (NCC) – green line, Structural Similarity Index (SSIM) – violet line, Relative Sum of Squared Differences (rel SSD) – yellow line, Normalized Gradient Field (NGF) – red line, Dice coefficient – dotted pink line, F1 score – dotted dark red line, and Jacard score – dotted magenta line.

VALOR-Net for Segmentation and Registration of the Breast MRI Data

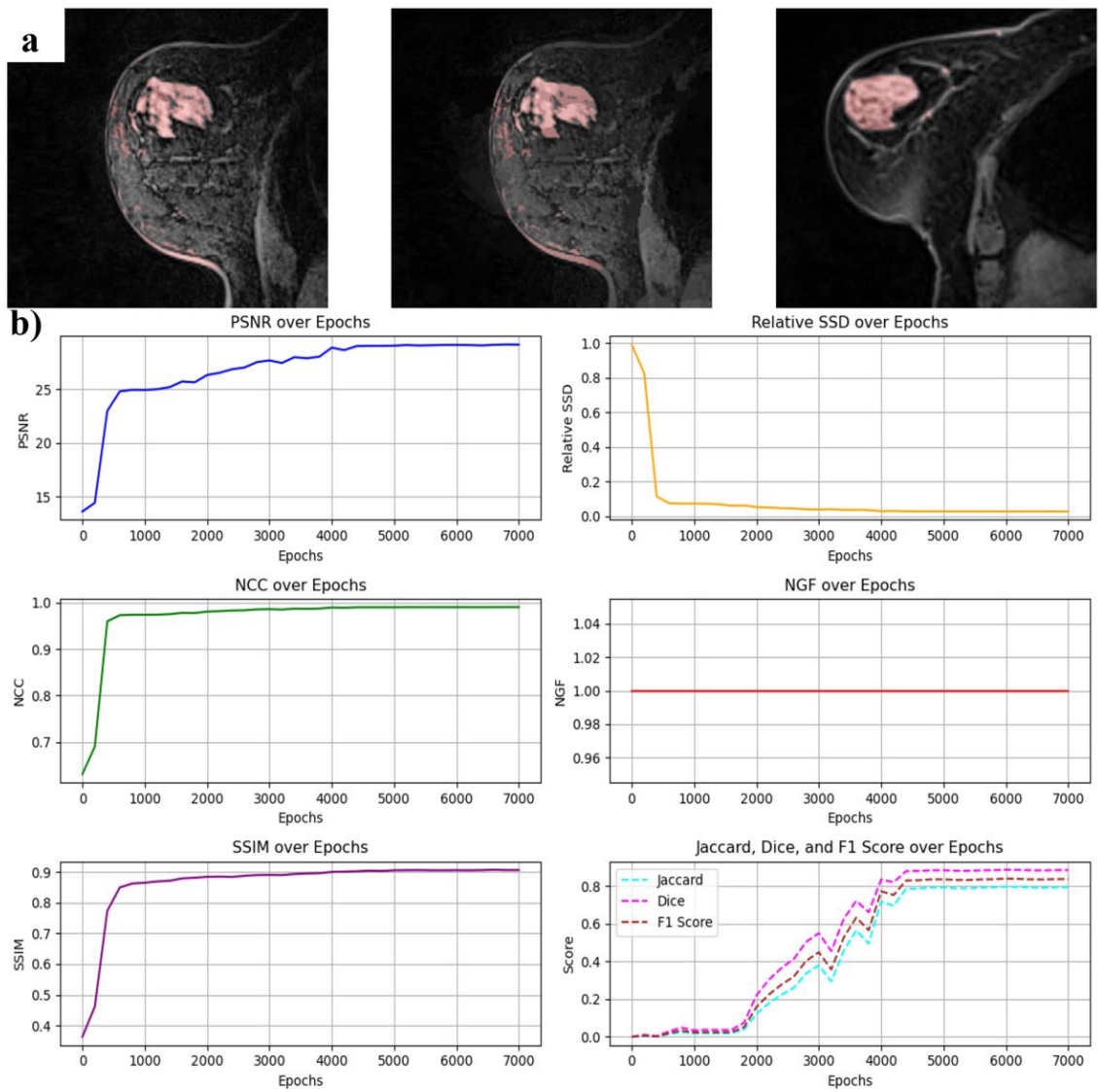

**Figure 6**: Tumor segmentation and registration results for subject 5: a) A 63-year-old female subject with invasive ductal carcinoma (IDC) tumor in her left breast. 3T MR image (left) shows heterogeneous tumor with a similar appearance on the 7T image (right) and on the transformed 3T image (middle). b) Quantitative metrics are presented as: Peak Signal-to-Noise Ratio (PSNR) – blue line, Normalized Cross-Correlation (NCC) – green line, Structural Similarity Index (SSIM) – violet line, Relative Sum of Squared Differences (rel SSD) – yellow line, Normalized Gradient Field (NGF) – red line, Dice coefficient – dotted pink line, F1 score – dotted dark red line, and Jacard score – dotted magenta line.

VALOR-Net for Segmentation and Registration of the Breast MRI Data

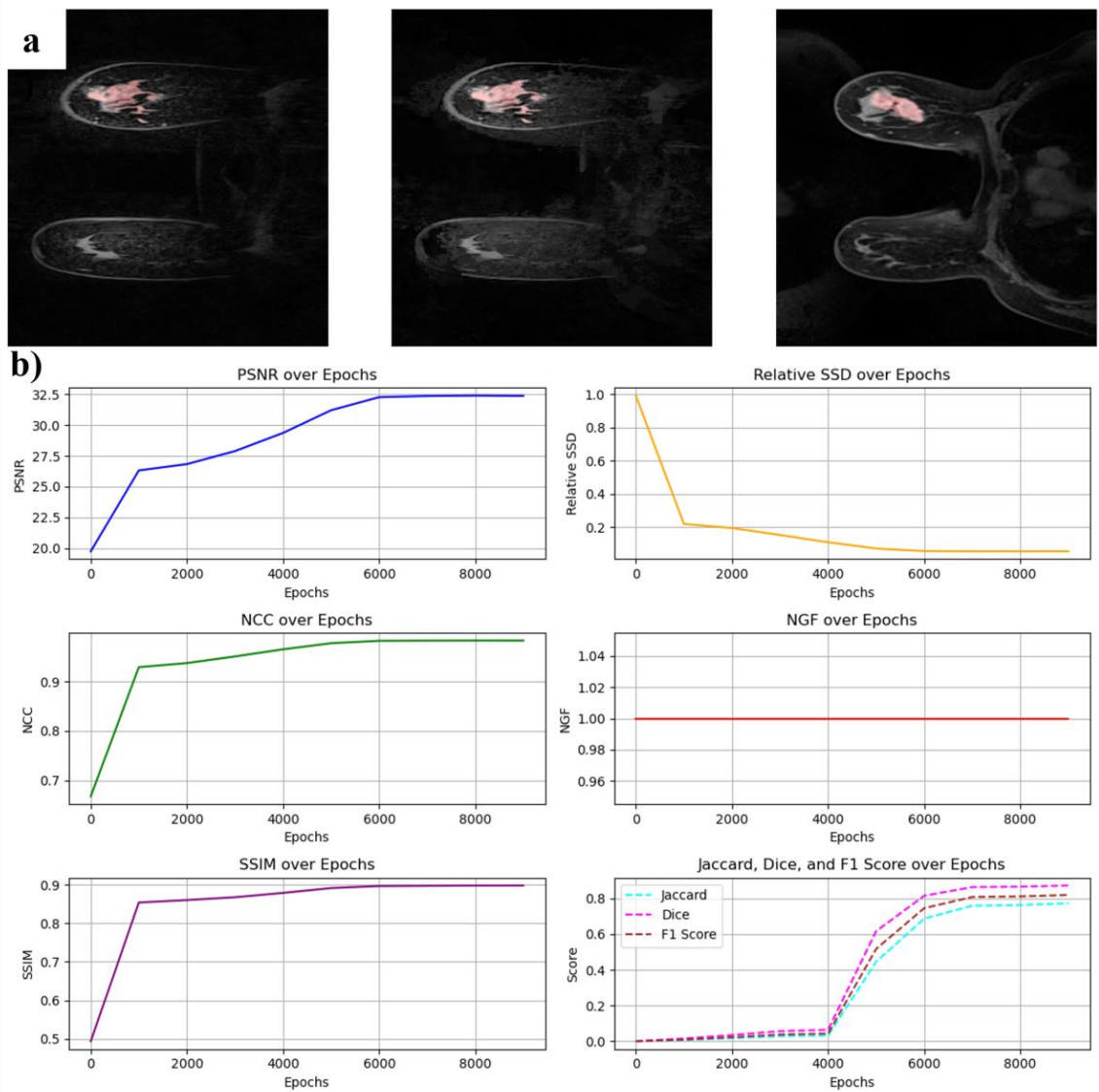

**Figure 7**: Tumor segmentation and registration results for subject 6: a) A 81-year-old female subject with invasive ductal carcinoma (IDC) in her left breast. 3T MR image (left) shows a large-size tumor with a similar appearance on the 7T image (right) and on the transformed 3T image (middle).   b) Quantitative metrics are presented as: Peak Signal-to-Noise Ratio (PSNR) – blue line, Normalized Cross-Correlation (NCC) – green line, Structural Similarity Index (SSIM) – violet line, Relative Sum of Squared Differences (rel SSD) – yellow line, Normalized Gradient Field (NGF) – red line, Dice coefficient – dotted pink line, F1 score – dotted dark red line, and Jacard score – dotted magenta line.

VALOR-Net for Segmentation and Registration of the Breast MRI Data

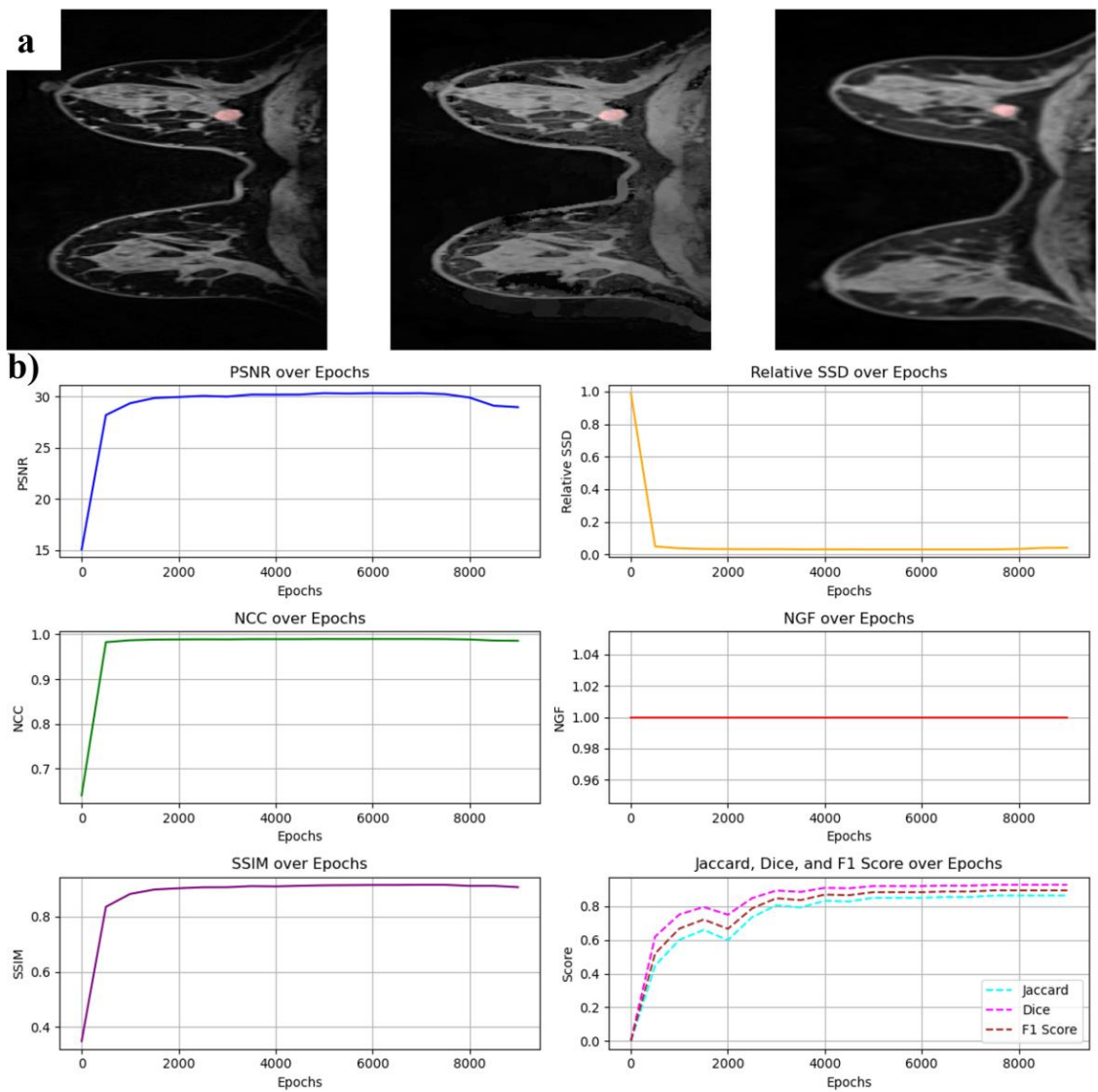

**Figure 8**: Tumor segmentation and registration results for subject 7: a) A 63-year-old female subject with a fibroadenoma in her right breast. 3T MR image (left) shows round-shaped tumor with regular margins also visible on the 7T image (right) and on the transformed 3T image (middle). b) Quantitative metrics are presented as: Peak Signal-to-Noise Ratio (PSNR) – blue line, Normalized Cross-Correlation (NCC) – green line, Structural Similarity Index (SSIM) – violet line, Relative Sum of Squared Differences (rel SSD) – yellow line, Normalized Gradient Field (NGF) – red line, Dice coefficient – dotted pink line, F1 score – dotted dark red line, and Jacard score – dotted magenta line.

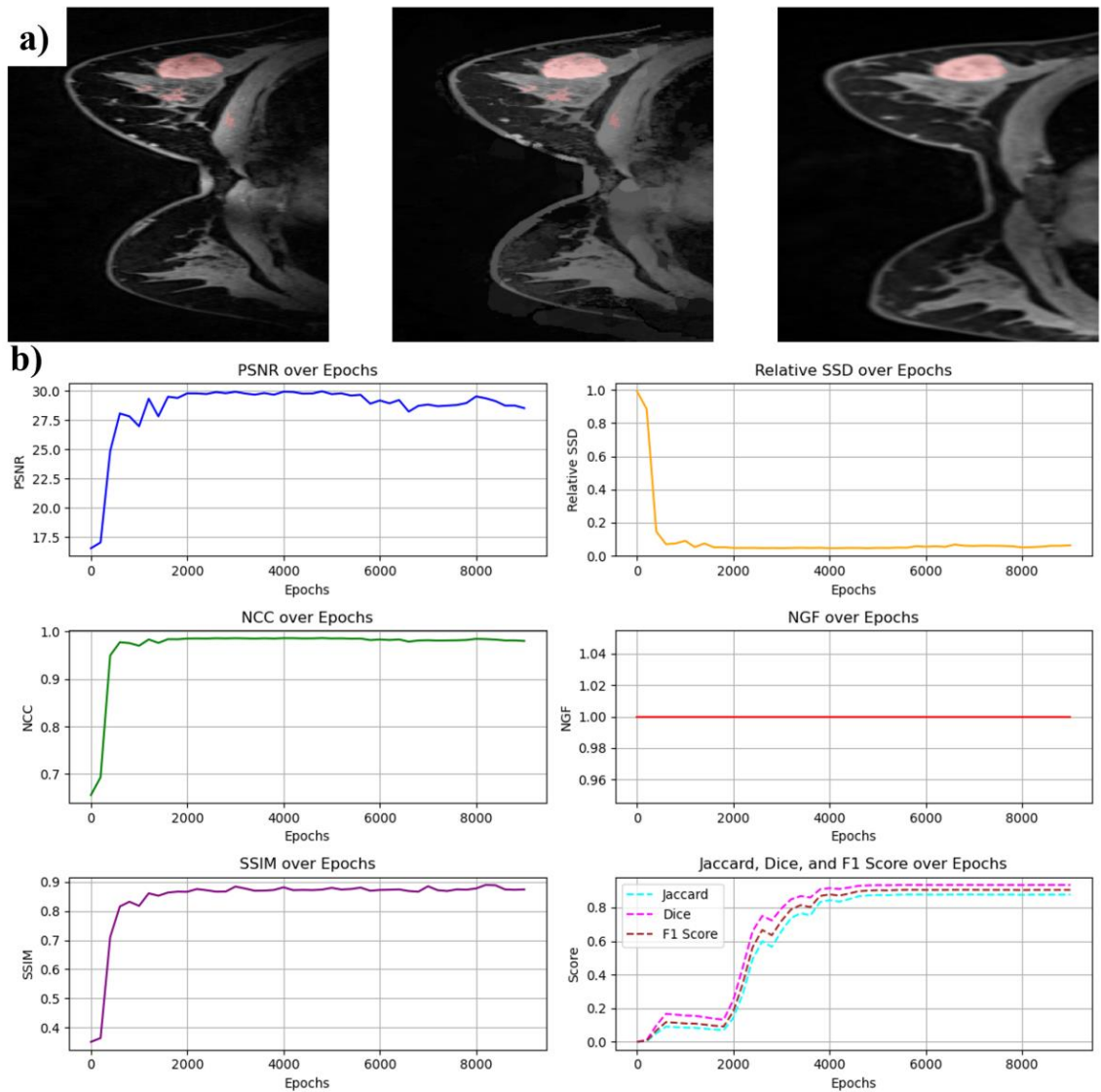

**Figure 9**: Tumor segmentation and registration results for subject 8: a) A 52-year-old female subject with a fibroadenoma in her left breast. 3T MR image (left) shows the 7T image demonstrating hyperintense tumor (right) and the transformed 3T image (middle).  b) Quantitative metrics are presented as: Peak Signal-to-Noise Ratio (PSNR) – blue line, Normalized Cross-Correlation (NCC) – green line, Structural Similarity Index (SSIM) – violet line, Relative Sum of Squared Differences (rel SSD) – yellow line, Normalized Gradient Field (NGF) – red line, Dice coefficient – dotted pink line, F1 score – dotted dark red line, and Jacard score – dotted magenta line.

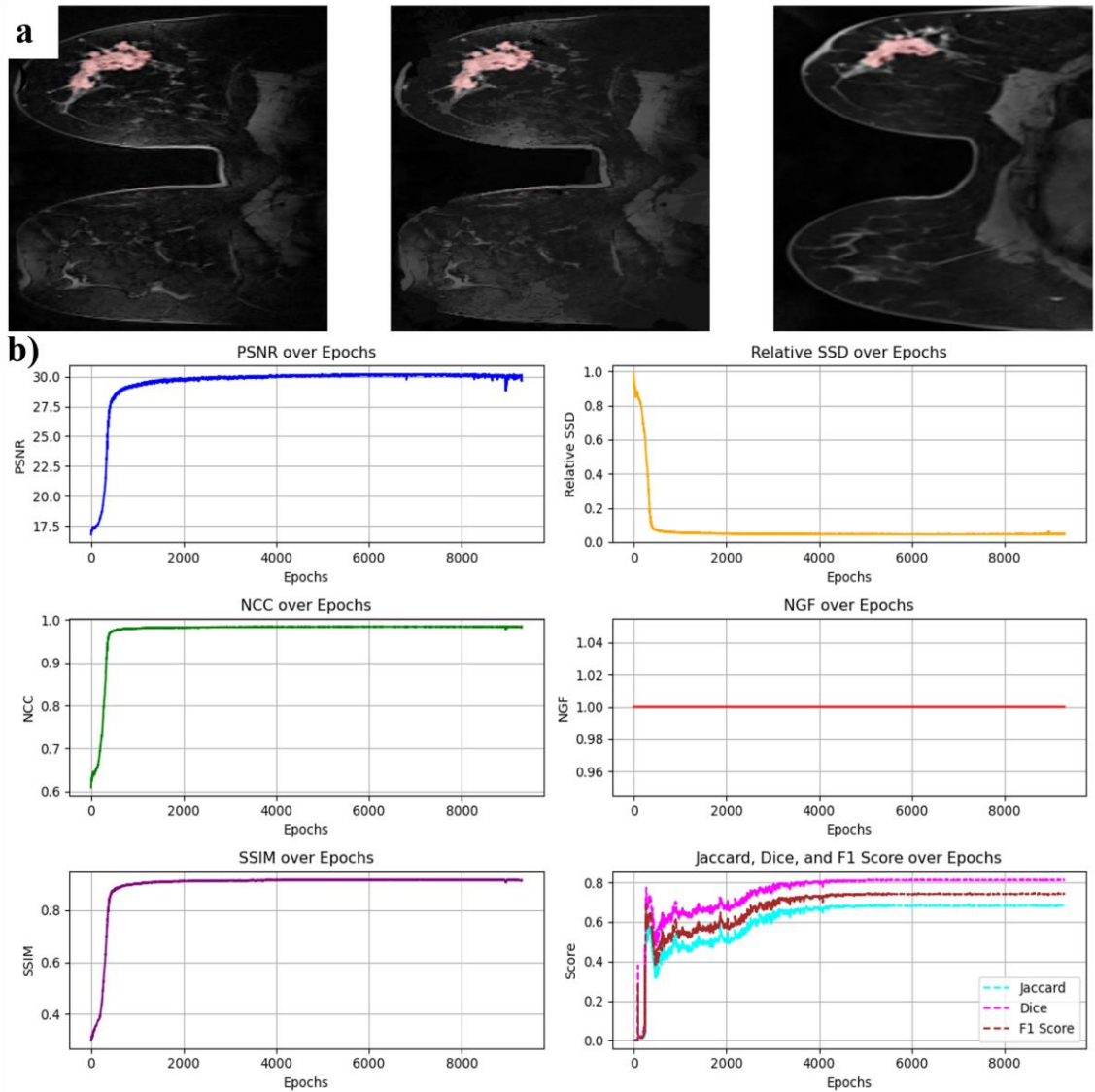

**Figure 10**: Tumor segmentation and registration results for subject 9: a) A 31-year-old female subject with an invasive ductal carcinoma (IDC) in her right breast. 3T MR image (left) shows heterogeneous tumor with spiculated margins also visible on the 7T image (right) and transformed 3T image (middle). b) Quantitative metrics are presented as: Peak Signal-to-Noise Ratio (PSNR) – blue line, Normalized Cross-Correlation (NCC) – green line, Structural Similarity Index (SSIM) – violet line, Relative Sum of Squared Differences (rel SSD) – yellow line, Normalized Gradient Field (NGF) – red line, Dice coefficient – dotted pink line, F1 score – dotted dark red line, and Jacard score – dotted magenta line.

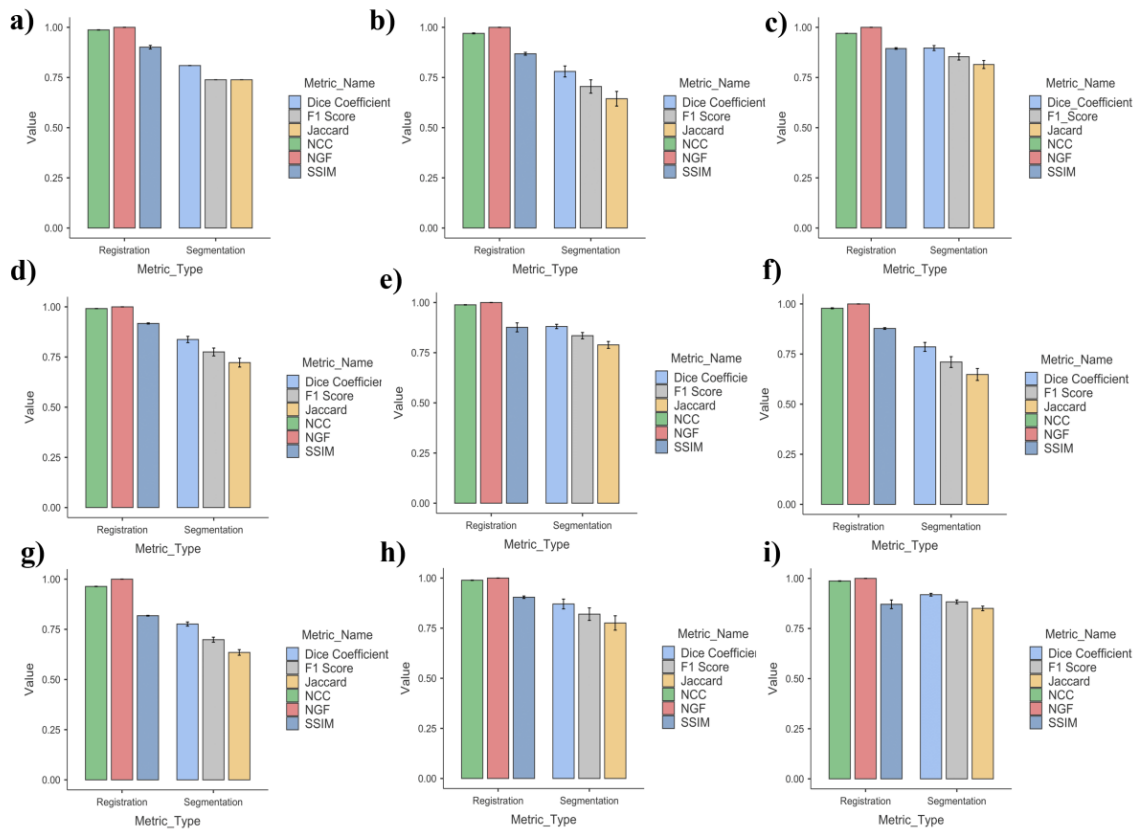

**Figure 11**: Aggregate tumor segmentation and registration performance across slices: a-i) calculated performance for each of the nine subjects.